\documentclass[twocolumn,showpacs,prb]{revtex4}
\newcommand{\figurewidth}{\columnwidth}
\usepackage{graphicx,amssymb}
\newcommand{\av}{_{\mathrm{av}}}

\begin{document}

\title{Large-scale Monte Carlo simulations of the isotropic three-dimensional
Heisenberg spin glass}

\author{L.~W.~Lee}
\email{leelikwe@u.washington.edu}
\affiliation{Department of Physics,
University of California,
Santa Cruz, California 95064}
\affiliation{Dept. of Mechanical Engineering,
University of Washington, Seattle, WA 98195}
\altaffiliation{Present address}
\author{A.~P.~Young}
\email{peter@bartok.ucsc.edu}
\homepage{http://bartok.ucsc.edu/peter}
\affiliation{Department of Physics,
University of California,
Santa Cruz, California 95064}

\date{\today}

\begin{abstract}
We study the Heisenberg spin glass by large-scale Monte Carlo
simulations for sizes up to $32^3$, down to temperatures below the
transition temperature claimed in earlier work. The data for the larger
sizes show more marginal behavior than that for the smaller sizes,
indicating the lower critical dimension is close to, and possibly equal
to three. We find that the spins and chiralities behave in a quite
similar manner.
\end{abstract}

\pacs{75.50.Lk, 75.40.Mg, 05.50.+q}
\maketitle

\section{Introduction}

Following the convincing numerical work of Ballesteros et
al.\cite{ballesteros:00} there has been little doubt that Ising spin
glasses in three dimensions have a finite temperature transition.  In
this paper we shall study a related model for which the existence of a
finite temperature transition is still controversial: the isotropic
Heisenberg spin glass, which is composed of classical spins with three
components.
Early
work \cite{mcmillan:85b,morris:86,olive:86} indicated a zero temperature
transition, or possibly a transition at a very low but non-zero
temperature.  However, following the pioneering work of
Villain\cite{villain:77b}, which emphasized the role of ``chiralities''
(Ising-like variables which describe the handedness of the non-collinear
spin structures), Kawamura and Tanemura\cite{kawamura:87} proposed, for
the XY case (which have two-component spins), that the spin glass transition
only occurs at $T_{SG} = 0$ and that a \textit{chiral}
glass transition occurs at
a finite temperature $T_{CG}$.  This scenario requires that spins and
chiralities decouple at long length scales. Kawamura and collaborators
subsequently proposed\cite{kawamura:98,hukushima:00,hukushima:05} that
this ``spin-chirality decoupling'' scenario also holds for Heisenberg
spin glasses.

However, the absence of a spin glass transition in Heisenberg spin
glasses has been challenged by Matsubara et
al.\cite{matsubara:01,endoh:01}, and Nakamura and
Endoh\cite{nakamura:02} who argued that the spins and chiralities order
at the same low but finite temperature. Recently Picco and
Ritort\cite{picco:05} also claimed a finite $T_{SG}$ and inferred that
probably $T_{SG} = T_{CG}$, though they did not investigate the
chiralities directly. In earlier work\cite{leeLW:03}, referred to as LY,
we studied spin and
chiral correlations on an equal footing, using the method of analysis
that was the most successful for the Ising spin
glass\cite{ballesteros:00}, namely finite-size scaling of the
correlation length. Considering a modest range of sizes, $L \le 12$,
we found that
the behavior of spins and chiralities was quite similar and they both
had a finite temperature transition, apparently at the same temperature.

However, quite recently Campos et al.\cite{campos:06} were able to study
larger sizes than LY, up to $L=32$. They agreed with LY that there is a
single transition at which both spins and chiralities order, but they
also found evidence for crossover, at the largest sizes, to a
``marginal'' behavior, reminiscent of that at the
Kosterlitz-Thouless-Berezenskii\cite{kosterlitz:73,berezinskii:70} (KTB)
transition in the two-dimensional XY ferromagnet where there is a finite
transition temperature $T_c$ but no long-range order for $T < T_c$. In
fact the region below $T_c$ is a \textit{line} of critical points in the
KTB theory.  If a line of critical points also exists in the
three-dimensional Heisenberg spin glass, then $d=3$ is the lower
critical dimension $d_l$, below which there is no transition. However,
given numerical uncertainties, Campos et al.~cannot rule out the
possibility that $d_l$ is slightly less than three, in which case there
\textit{is} spin glass order below $T_{SG}$.

Hukushima and
Kawamura\cite{hukushima:05} studied sizes up to $L=20$ and found more
marginal behavior when comparing $L=16$ and 20, than for the smaller
sizes. However, they argued that this effect is greater for the spin
correlation length than for the chiral correlation length, and hence concluded
that, while $T_{CG}$ is finite, $T_{SG}$ is zero or possibly non-zero
but less than $T_{CG}$, i.e.~there is spin-chirality decoupling.

In this paper we perform Monte Carlo simulations of the Heisenberg spin
glass, along the lines of LY, but for larger sizes. Our main motivation
is to investigate the claim of Campos et al.\cite{campos:06} that there
is a line of critical points for $T < T_{SG}$. In order to test whether
there is a critical \textit{line}, as proposed by Campos et al,
or the usual situation of a single critical
\textit{point} at $T_{SG}$, it is necessary to investigate the
behavior of the system \textit{below} the estimated $T_{SG}$. Campos et
al.~were not able to do this and the evidence for the critical
line was based on estimating corrections to scaling \textit{at}
$T_{SG}$. This is rather
\textit{indirect}, and perhaps not very reliable because the range of sizes
and quality of the data, are not sufficient to disentangle the various
corrections to scaling unambiguously. Very recently, the analysis of
Campos et al.~has been criticized by Campbell and
Kawamura\cite{campbell:07}. Here, we do not rely on
corrections to scaling but obtain data below $T_{SG}$ and so
\textit{directly} find evidence for more marginal behavior at larger
sizes. 

As we shall see, the transition temperature in the Heisenberg spin glass
is very low and so it is surprising to us that Campos et
al.~\cite{campos:06} did not use the technique of ``parallel
tempering''\cite{hukushima:96,marinari:98b}, which is the commonly used
approach to
speed up simulations of spin glasses, especially at very low
temperatures. Equilibrating lattices as large as $32^3$ is very
challenging, especially in the absence of parallel tempering. A second
motivation of our study is therefore
to use parallel tempering and to combine this
with a very useful equilibration test described in the next section, to
\textit{ensure} that the data are equilibrated.

From our results,
we see a strong crossover to much more marginal
behavior for sizes $L
\gtrsim 24$, in agreement with Campos et al.\cite{campos:06}. Whether
there is a KTB-like critical line as proposed by Campos et al.,
is, however, unclear. The lower critical
dimension, $d_l$ seems to be close to, or possibly equal to, three. The
behavior of the spins and chiralities is rather similar, and so,
in contrast to Kawamura and
collaborators\cite{hukushima:05,campbell:07}, we do \textit{not} feel
that our data supports spin-chirality decoupling.

\section{Model and analysis}
\label{sec:manda}

We use the standard Edwards-Anderson spin glass model
\begin{equation}
{\cal H} = -\sum_{\langle i, j \rangle} J_{ij} {\bf S}_i \cdot
{\bf S}_j,
\end{equation}
where the ${\bf S}_i$ are 3-component classical
vectors of unit length at the sites of a simple cubic lattice, and the
$J_{ij}$ are nearest neighbor interactions with a Gaussian distribution
with zero mean and standard deviation unity.  Periodic boundary
conditions are applied on lattices with $N=L^3$ spins. 

The spin glass order parameter, $q^{\mu\nu}({\bf k})$, at
wave vector ${\bf k}$, is defined to be
\begin{equation}
q^{\mu\nu}({\bf k}) = {1 \over N} \sum_i S_i^{\mu(1)} S_i^{\nu(2)}
e^{i {\bf k} \cdot {\bf R}_i},
\end{equation}
where $\mu$ and $\nu$ are spin components, and ``$(1)$'' and ``$(2)$''
denote two
identical copies of the system with the same interactions. From this we
determine the wave vector dependent
spin glass susceptibility $\chi_{SG}({\bf k})$ by
\begin{equation}
\chi_{SG}({\bf k}) = N \sum_{\mu,\nu} [\langle \left|q^{\mu\nu}({\bf
k})\right|^2 \rangle ]\av ,
\end{equation}
where $\langle \cdots \rangle$ denotes a thermal average and
$[\cdots ]\av$ denotes an average over disorder. The spin glass correlation
length
is then determined\cite{palassini:99b,ballesteros:00}, from 
\begin{equation}
\xi_L = {1 \over 2 \sin (k_\mathrm{min}/2)}
\left({\chi_{SG}(0) \over \chi_{SG}({\bf k}_\mathrm{min})} - 1\right)^{1/2},
\end{equation}
where ${\bf k}_\mathrm{min} = (2\pi/L)(1, 0, 0)$.

For the Heisenberg spin glass, Kawamura\cite{kawamura:98}
defines the local chirality
in terms of three spins on a line as follows:
\begin{equation}
\kappa_i^\mu = {\bf S}_{i+\hat{\mu}} \cdot {\bf S}_i \times {\bf
S}_{i-\hat{\mu}}.
\label{eq:chiral_heis}
\end{equation}
The chiral glass susceptibility is then given by
\begin{equation}
\label{chisg}
\chi_{CG}^\mu({\bf k}) =  N [\langle \left| q_{c}^\mu({\bf k})\right|^2
\rangle ]\av ,
\end{equation}
where the chiral overlap $q_{c}^\mu({\bf k})$ is given by
\begin{equation}
\label{qc}
q_{c}^\mu({\bf k}) = {1 \over N} \sum_i  \kappa_i^{\mu(1)} \kappa_i^{\mu(2)}
e^{i {\bf k} \cdot {\bf R}_i}.
\end{equation}
We define
the chiral correlation lengths $\xi^\mu_{c,L}$ by
\begin{equation}
\label{xi_c}
\xi^\mu_{c,L} = {1 \over 2 \sin (k_\mathrm{min}/2)}
\left({\chi_{CG}(0) \over \chi_{CG}^\mu({\bf k}_\mathrm{min})} - 1\right)^{1/2},
\end{equation}
in which $\chi_{CG}({\bf k}=0)$ is independent of $\mu$.
Note that $\xi^\mu_{c,L}$ will, in general, be different for $\hat{\mu}$ along
${\bf k_\mathrm{min}}$ (the $\hat{x}$ direction)
and perpendicular to ${\bf k}$, though we expect that this difference
will vanish for large sizes. We 
denote these two lengths by $\xi^\parallel_{c,L}$ and $\xi^\perp_{c, L}$
respectively.

To equilibrate the system in as small a number of sweeps as
possible, with the minimum amount of CPU time we perform three types of
Monte Carlo move:
\begin{enumerate}
\item
``Microcanonical'' sweeps\cite{alonso:96},\\
(also known as ``over-relaxation'' sweeps).
We sweep sequentially through the lattice, and, at
each site, compute the local field on the spin, $\mathbf{H}_i = \sum_j J_{ij}
\mathbf{S}_j$. The new value for the spin on site $i$ is taken to be
its old value reflected about $\mathbf{H}$, i.e.
\begin{equation}
\mathbf{S}'_i = -\mathbf{S}_i + 2\, {\mathbf{S}_i \cdot \mathbf{H}_i \over
H_i^2}\, \mathbf{H}_i \, ,
\label{reflect}
\end{equation}
see Fig,~\ref{micro}. These sweeps are microcanonical
because they preserve energy. They are very fast because the
operations are simple and no random numbers are needed.  For reasons
that are not fully understood, it also seems that they ``stir up'' the
spin configuration very efficiently\cite{campos:06} and the system
equilibrates faster than if one only uses ``heatbath'' updates,
described next.

\begin{figure}[!tbp]
\begin{center}
\includegraphics[width=4cm]{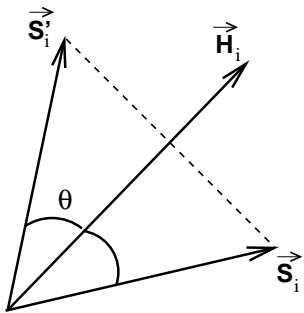}
\caption{
$\mathbf{H}_i$ is the local field on site $i$ due to its neighbors.
The spin at $i$ is
initially in direction $\mathbf{S}_i$. In a microcanonical
(over-relaxation) move, the spin is reflected about $\mathbf{H}_i$
according to Eq.~(\ref{reflect}), and so
ends up in direction $\mathbf{S}'_i$.
\label{micro}
}
\end{center}
\end{figure}

\item ``Heatbath'' sweeps\cite{olive:86}.\\ Since the microcanonical
sweeps conserve energy they do not equilibrate the system. We therefore
also include some heatbath sweeps since these do change the energy,
typically doing one after every 10 microcanonical sweeps.  As for the
microcanonical case, we sweep sequentially through the lattice.
Referring to Fig.~\ref{micro}, we take the direction of $\mathbf{H}_i$,
the local field on site, $i$ to be the
polar axis, denote the polar angle as $\theta$, and define the azimuthal
angle $\phi$ such that $\phi = 0$ for the old spin direction. The new
spin direction $\mathbf{S}'_i$ is characterized by angles $\theta$ and $\phi$,
relative to $\mathbf{H}_i$, as
follows. The energy does not depend on the azimuthal angle, and so $\phi
= 2 \pi r_1$, where $r_1$ is a random number chosen uniformly between 0
and 1. The polar angle is chosen such that, after the move, the spin is
in local equilibrium with respect to the local field, i.e.~if $x =
\cos\theta$, then 
\begin{equation}
P(x) = {\beta H_i \over 2 \sinh \beta H_i} \, e^{\beta H_i x} ,
\end{equation}
where $\beta = 1/ T$.
To determine $x$ with
this probability, the procedure\cite{olive:86} is to equate the
cumulative distribution
\begin{equation}
Q(x) = \int_{-1}^x P(x')\, dx' = {e^{\beta H_i x} - e^{-\beta H_i} \over
e^{\beta H_i } - e^{-\beta H_i} } , 
\end{equation}
to a second random number $r_2$ (also with a uniform distribution
between 0 and 1) so
\begin{equation}
x =  {1 \over \beta H_i} \, \ln \left[1+r_2 \left(e^{2\beta H_i} -
1\right)\right]
-1 \, .
\end{equation}
It is then necessary to convert the new spin direction
$\mathbf{S}'_i$ back to cartesian
coordinates. Denoting the polar and azimuthal angles of $\mathbf{H}$ by
$\theta_H$ and $\phi_H$ relative the cartesian reference frame,
and remembering that $\theta$ and $\phi$ are
relative to $\mathbf{H}$, we have\cite{comment1}
\begin{eqnarray}
S_x^\prime & = & S_x^{\prime\prime} \cos\phi_H -
S_y^{\prime\prime} \sin\phi_H \, ,\\
S_y^\prime & = & S_x^{\prime\prime} \sin\phi_H +
S_y^{\prime\prime} \cos\phi_H \, ,\\
S_z^\prime & = & \cos\theta \cos\theta_H - \sin\theta \sin\theta_H
\cos\phi \, ,
\end{eqnarray}
where
\begin{eqnarray}
S_x^{\prime\prime} & = & \cos\theta \sin\theta_H +
\sin\theta \cos\theta_H \cos\phi\, , \\
S_y^{\prime\prime} & = & \sin\theta \sin\phi \, .
\end{eqnarray}
We see that the calculations in the heatbath moves are quite
involved, which is why we do mainly microcanonical moves, just including
\textit{some} heatbath moves to change the energy and thereby ensure the
algorithm is ergodic. Note, though, that the
acceptance probability for the heat bath moves, and also for the
microcanonical moves, is unity, so \textit{no moves are wasted}.

\item
``Parallel tempering'' sweeps.\\
At low temperatures spin glasses are easily trapped in minima (valleys)
of the
free energy. In order to ensure that the system visits different minima
with the correct Boltzmann weight during the time of the simulation we
use the method of parallel tempering\cite{hukushima:96,marinari:98b}.
One takes $N_T$ copies of the system with the same bonds but at
a range of
different temperatures.
The minimum temperature, $T_{\rm min} \equiv T_1$,
is the low temperature where one
wants to investigate the system (below $T_{SG}$ in our case), and the
maximum,
$T_{\rm max} \equiv T_{N_T}$,
is high enough that the
the system equilibrates very fast
(well above $T_{SG}$ in our case). A parallel tempering sweep consists
of swapping the temperatures of the spin configurations at a pair of
neighboring temperatures, $T_i$ and $T_{i+1}$, for $i = 1, 2, \cdots ,
T_{N_T - 1}$ with a probability that satisfies the detailed balance
condition. The Metropolis probability for this is\cite{hukushima:96}
\begin{equation}
\quad P(T\ \mbox{swap}) = \left\{
\begin{array}{ll}
\exp(-\Delta \beta \, \Delta E),  & (\mbox{if} \ \Delta \beta \, \Delta E
> 0), \\
1, & (\mbox{otherwise}), 
\end{array}
\right.
\label{PTswap}
\end{equation}
where $\Delta \beta= 1/T_i - 1/T_{i+1}$ and $\Delta E = E_i - E_{i+1}$,
in which $E_i$ is the energy of the copy at temperature $T_i$. In this
way, a given set of spins (i.e.~a copy)
performs a random walk in temperature space.
Suppose that at some time in the simulation
a copy is trapped in a valley at low-$T$. Later on
it will reach a high temperature where it randomizes quickly, so that
when, still later, it is again at a low temperature, there is no reason
for it to be in the same valley that it was in before. We do one
sweep of temperature swaps after every ten microcanonical sweeps.

Table \ref{simparams} gives the parameters of the simulations. It will
be seen that the number of temperatures is very large, and it is
instructive
to discuss the reason for this. The difference between two neighboring
temperatures, $\Delta T$, must be sufficiently small that there is an
overlap between the energy distributions at those temperatures, so
there is a reasonable
probability that the same spin configuration occurs for 
both temperatures. Otherwise, the probability
for accepting the temperature swap, Eq.~(\ref{PTswap}), will be very
small. Relating the width of the temperature distribution to the
specific heat in the normal way, one finds that
\begin{equation}
{\delta T \over T} \lesssim {1 \over \sqrt{C \, N}} \, ,
\end{equation}
where $C$ is the specific heat per spin. For the Heisenberg spin glass
$C$ tends to a constant at low-$T$ because of the Gaussian (spinwave)
fluctuations about the local equilibrium positions. Hence, for a given
size, we choose temperatures
which decrease in a geometric manner.  The required
number of temperatures to cover a fixed range of $T$ increases
proportional to $\sqrt{N}$, and so, for large $N$, many copies are needed to
cover even a factor of 2 in $T$. 
However, for the
Ising spin glass with Gaussian interactions, the specific heat tends to zero
as $T \to 0$ (presumably linearly in $T$). Hence much bigger steps in $T$ can
be taken at low-$T$ than for the Heisenberg case, leading to fewer
temperatures being needed.

\end{enumerate}

\begin{table}
\caption{
Parameters of the simulations.  $N_{\rm samp}$ is the number of samples,
$N_{\rm equil}$ is the number of microcanonical Monte Carlo sweeps for
equilibration for each of the $2 N_T$ replicas for a single sample, and
$N_{\rm meas}$ is the number of microcanonical sweeps for measurement.
The number of heatbath sweeps is equal to
10\% of the number of
microcanonical sweeps.  $T_{\rm min}$ and $T_{\rm max}$ are the
lowest and highest temperatures simulated,
and $N_T$ is the number of temperatures
used in the parallel tempering.
\label{simparams}
}
\begin{tabular*}{\columnwidth}{@{\extracolsep{\fill}} r r r r r r r } 
\hline
\hline
$L$  & $N_{\rm samp} $ & $N_{\rm equil}$ & $N_{\rm meas}$ & $T_{\rm min}$ &
$T_{\rm max}$ & $N_{T}$  \\ 
\hline
 4 &  500 & $5   \times 10^3$ & $10^4$             & 0.0400 & 0.96 &  40 \\
 6 &  500 & $6   \times 10^4$ & $1.2  \times 10^5$ & 0.0400 & 0.96 &  40 \\
 8 &  536 & $8   \times 10^4$ & $1.6  \times 10^5$ & 0.0400 & 0.96 &  57 \\
12 &  204 & $8   \times 10^5$ & $1.6  \times 10^6$ & 0.0400 & 0.61 &  88 \\
16 &  202 & $8   \times 10^5$ & $1.6  \times 10^6$ & 0.1015 & 0.49 &  73 \\ 
24 &  160 & $3.2 \times 10^5$ & $6.4  \times 10^5$ & 0.1200 & 0.49 & 122 \\ 
32 &   56 & $9.6 \times 10^5$ & $1.28 \times 10^6$ & 0.1210 & 0.40 & 120 \\ 
\hline
\hline
\end{tabular*}
\end{table}

To test for
equilibration\cite{katzgraber:01} we require that data satisfy the 
relation\cite{katzgraber:01c}
\begin{equation}
q_l - q_s 
=  {2\over z}\, T\, U,
\label{equiltest}
\end{equation}
which is valid for a Gaussian bond distribution. Here
$U = - [\sum_{\langle i, j \rangle} J_{ij} \langle {\bf S}_i \cdot
{\bf S}_j \rangle ]\av $ is
the average energy per spin, $q_l = (1/N_b)\sum_{\langle i, j \rangle}
[ \langle
{\bf S}_i \cdot {\bf S}_j \rangle^2]\av$ is the ``link overlap'', $q_s =
(1/N_b)\sum_{\langle i, j \rangle}[\langle ({\bf S}_i \cdot {\bf S}_j)^2
\rangle]\av$, $N_b = (z/2)N$ is the number of nearest neighbor bonds, and
$z\ (=6\ \mbox{here})$ is the lattice coordination number. Equation
(\ref{equiltest}) is easily derived by integrating by parts the
expression for the average energy with respect to $J_{ij}$, noting that the
average $[\cdots]\av$ is over a Gaussian function of the
$J_{ij}$'s.

\begin{figure}[!tbp]
\begin{center}
\includegraphics[width=\figurewidth]{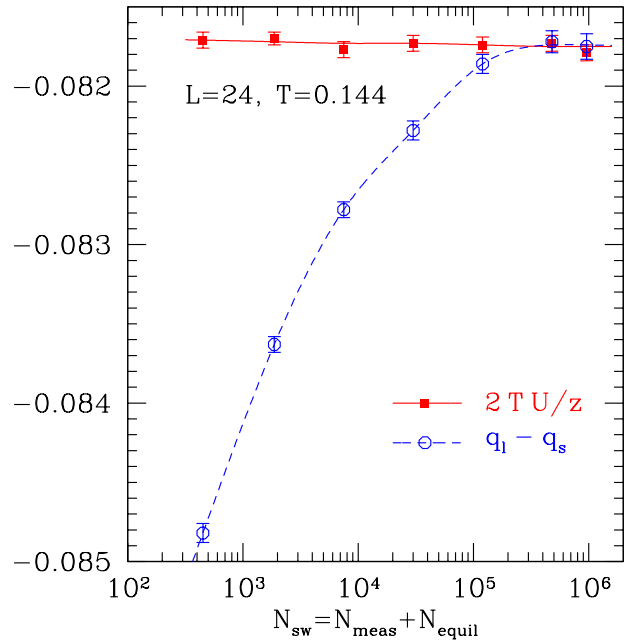}
\end{center}
\caption{(Color online)
Equilibration plot, testing Eq.~(\ref{equiltest})
for $L=24$ at the $T=0.144$. It is seen that
the data for $2T U/z$ come together at about $3 \times 10^5$ total
sweeps (equilibration plus measurement) and then stay at their common value
indicating that equilibration has been achieved. The lines are guides to
the eye. It is seen that the
energy comes close to its equilibrium value very quickly,
whereas $q_l - q_s$ takes much longer.
\label{equil}
}
\end{figure}

\begin{figure}[!tbp]
\begin{center}
\includegraphics[width=\figurewidth]{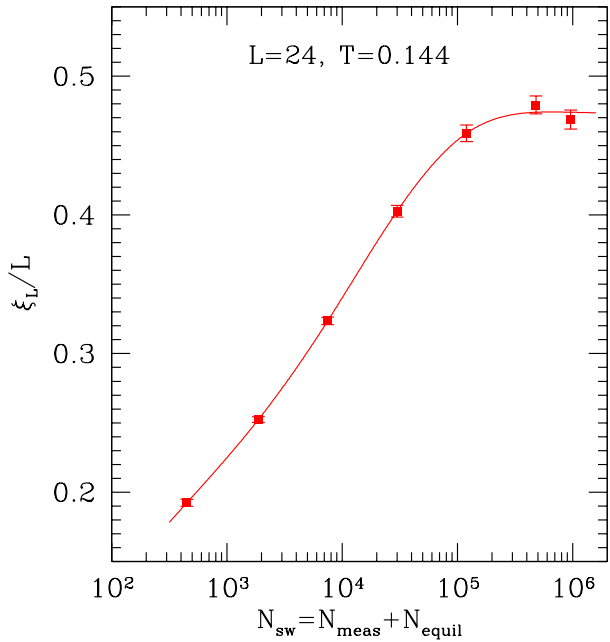}
\end{center}
\caption{(Color online)
A plot of $\xi_L/L$ as a function of the total number of sweeps for
$L=24$ at $T=0.144$. It is seen the data flattens
off at around $3 \times 10^5$ sweeps, the value where the two sets of
data in Fig.~\ref{equil} start to agree. This indicates that when the
data in Fig.~\ref{equil} agree within high precision, i.e.~when
Eq.~(\ref{equiltest}) is satisfied, the correlation length has reached
its equilibrium value. The line is a guide to the eye.
\label{equil_xi} 
}
\end{figure}

\begin{figure}[!tbp]
\begin{center}
\includegraphics[width=\figurewidth]{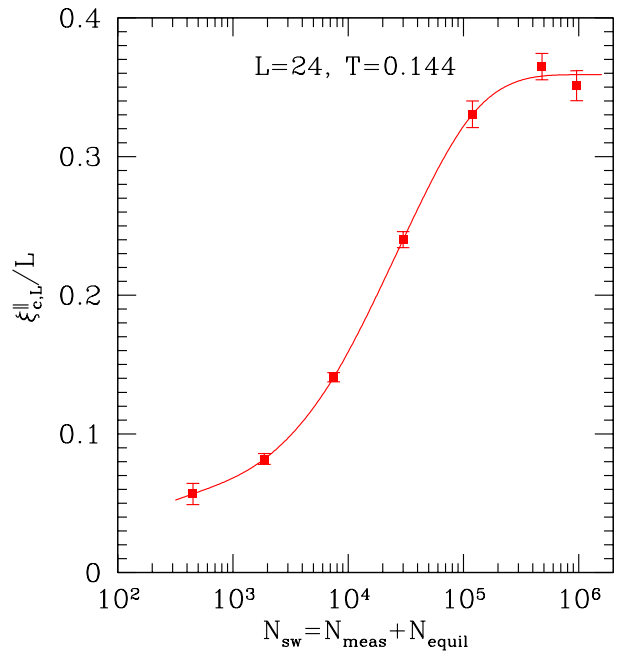}
\end{center}
\caption{(Color online)
A plot of $\xi^\parallel_{c,L}/L$
as a function of the total number of sweeps for
$L=24$ at $T=0.144$. As for Fig.~\ref{equil_xi} 
it is seen the data flattens
off at around $3 \times 10^5$ sweeps, the value where the two sets of
data in Fig.~\ref{equil} start to agree. The line is a guide to the
eye.
\label{equil_xicx} 
}
\end{figure}

The spins are initialized in random directions so the energy, the RHS of
Eq.~(\ref{equiltest}), is initially close to zero and decreases,
presumably monotonically, to its equilibrium value as the length of the
simulation increases. Hence the RHS of Eq.~(\ref{equiltest}) will be too
\textit{large} if the simulation is too short to equilibrate the system.
On the other hand, the LHS of Eq.~(\ref{equiltest}), $q_l$ will be too
\textit{small} if the simulation is too short because it starts off
close to zero and then increases with MC time as the two replicas start
to find the same local minima. The quantity $q_s$ will be less dependent
on Monte Carlo time than $q_l$ since it is a local variable for a single
replica.  (For the Ising case it is just a constant.) Hence if the
simulation is too short the LHS of Eq.~(\ref{equiltest}) will be too
low. In other words, the two sides of Eq.~(\ref{equiltest}) approach the
common equilibrium value from \textit{opposite directions} as the length
of the simulation increases.
Only if Eq.~(\ref{equiltest}) is satisfied within small error
bars do we accept the results of a simulation.

Figure \ref{equil} shows a test to verify that Eq.~(\ref{equiltest}) is
satisfied at long times. For the parameters used, $L=24, T=0.144$, this
occurs when the total number of sweeps ($N_{\rm sw}=N_{\rm equil}+N_{\rm meas}$)
is about $3 \times 10^5$. Figures
\ref{equil_xi} and \ref{equil_xicx} show that the spin
and chiral correlation lengths appear to become independent of $N_{\rm sw}$,
and hence are presumably equilibrated, when $N_{\rm sw}$ is larger than this
\textit{same} value. Hence, it appears that when Eq.~(\ref{equiltest}) is
satisfied to high precision, the data for the correlation lengths is
equilibrated.

With the number of sweeps shown in Table \ref{simparams},
Eq.~(\ref{equiltest}) was satisfied for all sizes and temperatures. The
error bars are made sufficiently small by averaging over a large number
of samples. We are simulating system sizes which are very large by spin
glass standards (up to $N=32^3$), so it is crucial to have a stringent
test like this for equilibration.

Since $\xi_L/L$ is dimensionless it has the finite size scaling
form\cite{ballesteros:00,palassini:99,lee:03}
\begin{equation}
{\xi_L \over L} = \widetilde{X}\left(L^{1/\nu}(T - T_{SG})\right) ,
\label{eq:fss}
\end{equation}
where $\nu$ is the correlation length exponent.  Note that there is no
power of $L$ multiplying the scaling function $\widetilde{X}$, as there
would be for a quantity with dimensions.  There are analogous
expressions for the chiral correlation lengths.  From Eq.~(\ref{eq:fss})
it follows that the data for $\xi_L/L$ for
different sizes come together at
$T=T_{SG}$. In addition, they are also expected to splay out again on
the low-$T$ side if there is spin glass order below $T_{SG}$. In a
marginal situation, with a line of critical points as in the KTB
transition, the data for different sizes would come together at $T_{SG}$
and then stick together at lower $T$, see, for example, Fig.~3 of
Ref.~\onlinecite{ballesteros:00}.

\begin{figure}[!tbp]
\begin{center}
\includegraphics[width=\figurewidth]{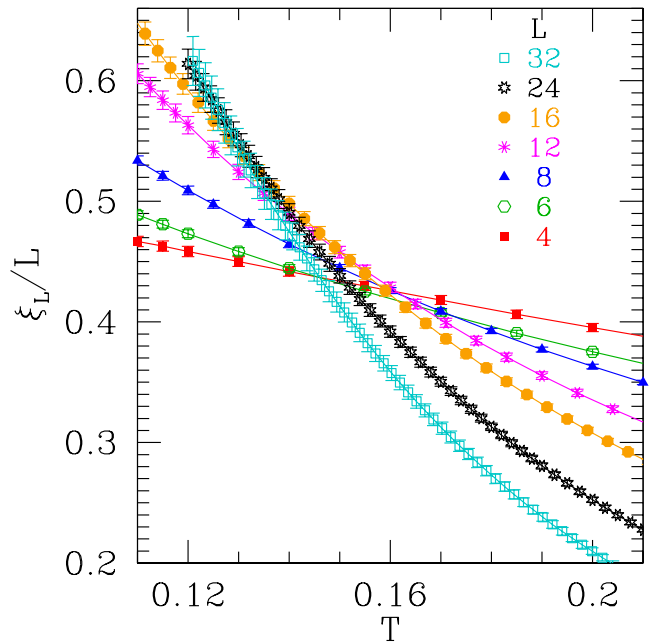}
\end{center}
\caption{(Color online)
Data for $\xi_L/L$,
the spin glass correlation length divided by system size,
as a function of $T$ for different system sizes. Note that there are
very many data points for the larger sizes. This is because a large
number of temperatures are needed for the parallel tempering algorithm,
as discussed in Sec.~\ref{sec:manda}.
\label{xi_L} 
}
\end{figure}

\begin{figure}[!tbp]
\begin{center}
\includegraphics[width=\figurewidth]{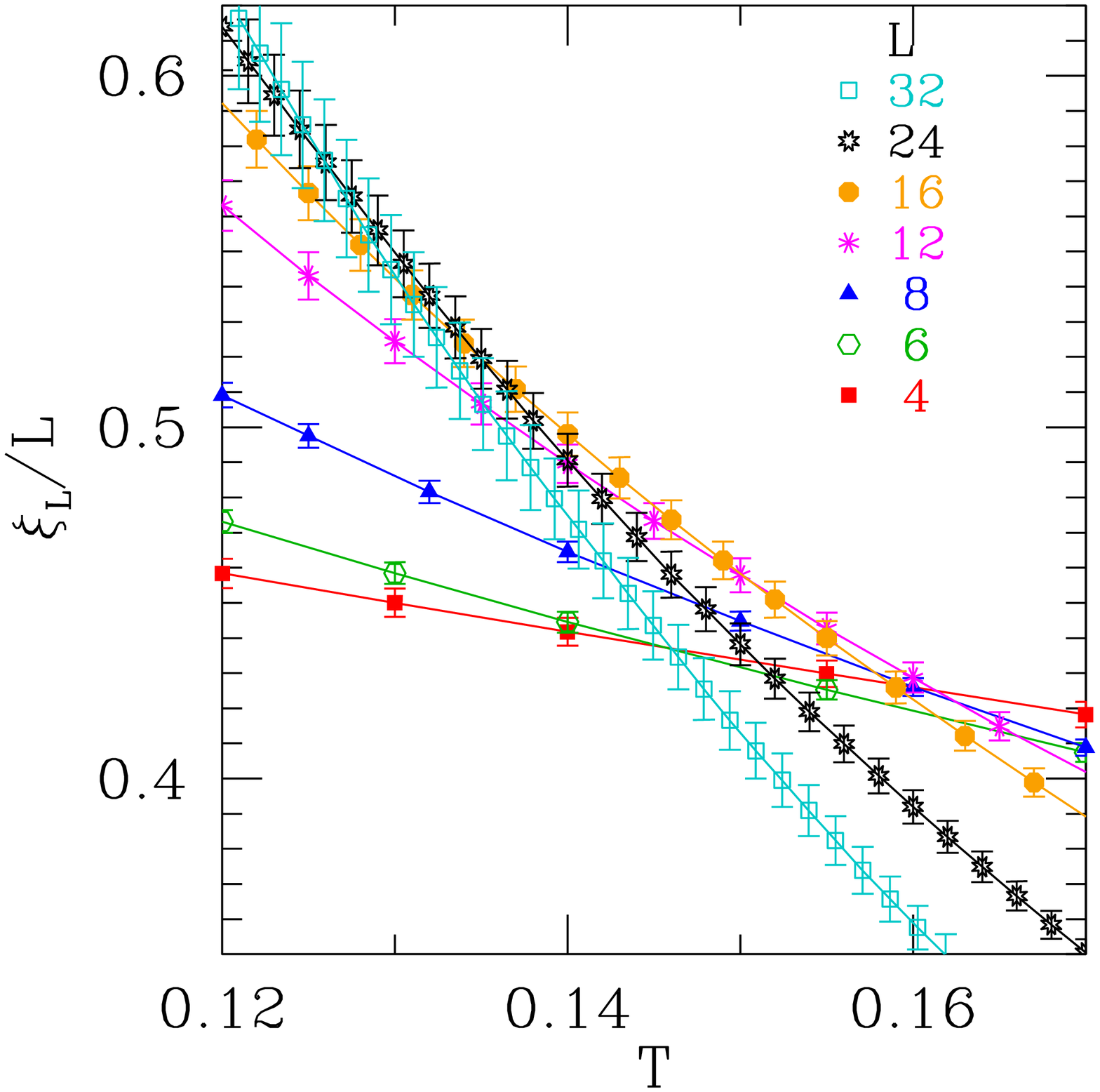}
\end{center}
\caption{(Color online)
Enlarged view of a region of Fig.~\ref{xi_L}.
\label{xi_L_enlarge} 
}
\end{figure}

\begin{figure}[!tbp]
\begin{center}
\includegraphics[width=\figurewidth]{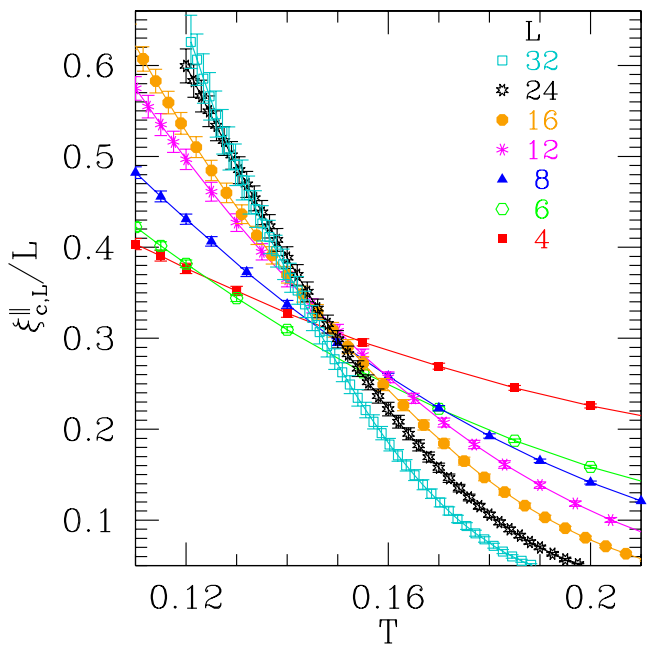}
\end{center}
\caption{(Color online)
Data for $\xi^\parallel_{c,L}/L$,
the ``parallel'' chiral glass correlation length divided by system size,
as a function of $T$ for different system sizes.
\label{xi_chiralx_L} 
}
\end{figure}

\begin{figure}[!tbp]
\begin{center}
\includegraphics[width=\figurewidth]{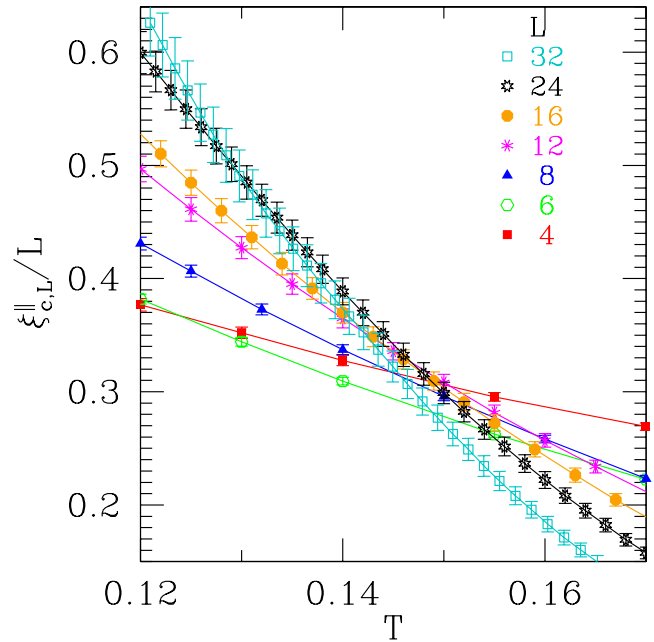}
\end{center}
\caption{(Color online)
Enlarged view of a region of Fig.~\ref{xi_chiralx_L}.
\label{xi_chiralx_L_enlarge} 
}
\end{figure}

\section{Results}
We studied sizes from $L=4$ to $L=32$, as shown in Table
\ref{simparams}. The CPU time involved to get this data is about
15 Mac G5 CPU years.

We start with the data for $\xi_L/L$ shown in Fig.~\ref{xi_L}. As was found
earlier by LY, the smaller sizes show a clear intersection, and also
splay out at lower temperatures which would indicate spin glass order.
However, the situation for the largest sizes is less clearcut, with the
$L=32$ and $24$ data only coming together at a somewhat lower
temperature than the temperature where the $L \le 16$ data intersect.
Furthermore the data for the largest sizes at the lowest temperatures
does not depend strongly on size, indicating close to marginal behavior.
This is qualitatively in agreement with Campos et al.\cite{campos:06}.
An enlarged view of the important region is shown in
Fig.~\ref{xi_L_enlarge}.

Data for the parallel chiral glass correlation length is shown in
Fig.~\ref{xi_chiralx_L}. The main features are the same as found for the 
spin glass correlation length in Fig.~\ref{xi_L}. The smaller sizes show
clear intersections, but the $L=32$ data lies lower, though perhaps not
quite to the same extent as in Fig.~\ref{xi_L}. As for the spin glass
correlation length, the two larger sizes $L=32$ and 24 only come
together at a lower temperature than the temperature (range) where the
smaller sizes intersect, and do not splay out at still lower
temperatures, at least in the range of $T$ studied and within the error
bars.  An enlarged view is shown in Fig.~\ref{xi_chiralx_L_enlarge}. In
our view, the data in Figs.~\ref{xi_L} and \ref{xi_chiralx_L} are not
very different; in particular, in both figures, the data for the largest
\textit{two} sizes merge at about the same temperature and stick together at
lower temperatures.  If we look at the largest \textit{three} sizes
there \textit{is} a somewhat greater tendency for the chiral data to
splay out at low-$T$. Clearly larger sizes are needed to be sure of the
trend in the thermodynamic limit.

With this data it is difficult to come to a firm conclusion about the
nature of the transition. Clearly the smaller sizes have large
corrections to scaling behavior, and so, at best, only the two largest
sizes are in the asymptotic scaling region. The data is consistent with
a line of critical points analogous to that in the KTB transition, as
proposed by Campos et al., in which the data for large sizes would merge
at a single temperature and remain independent of size at lower
temperatures. This is a scenario in which the lower critical dimension $d_l$
is equal to three.  Another scenario consistent with our data,
in which $d_l$ is also equal to $3$, is that
$T_{SG}=T_{CG}=0$ but with an exponential divergence of the correlation
lengths at $T=0$. In that case, it is likely that the data for large
sizes would join a common curve, but the temperature at which the common
curve is joined would decrease as the system size increased. This scenario is
found in the two-dimensional Heisenberg ferromagnet, whereas the
critical line of the KTB theory is found in the two-dimensional
XY-ferromagnet. Our data is also consistent with the
possibility that, for the large sizes, the
curves weakly intersect, implying a finite $T_{SG}$ and a lower critical
dimension slightly \textit{less} than three.

Unfortunately, because of the crossover effects in the data, it does not
appear possible to give a meaningful estimate of the critical exponents,
$\nu$ and $\eta$.

\section{Conclusions}

In agreement with Campos et al.\cite{campos:06} and 
Hukushima and Kawamura\cite{hukushima:05}, we find a crossover in
behavior for system sizes larger than about 16. Results from the larger
sizes indicate a more marginal behavior than those from the smaller
sizes. Unlike Kawamura and collaborators\cite{hukushima:05,campbell:07},
we do not find
that this effect is very different for the spins and chiralities.
Compare, for example, Figs.~\ref{xi_L} and \ref{xi_chiralx_L}. However,
larger
sizes would be needed to \textit{confirm} that the asymptotic behaviors of the
spin and chiral glass correlation lengths are indeed similar.
Our data
extends up to $L=32$, somewhat larger than the sizes ($L \le 20$) studied by
Hukushima and Kawamura.
Our range of sizes is the same as that of Campos et al., but we
are able to go down to lower temperature, in particular \textit{below}
the putative spin glass transition temperature. Whereas Campos et al.
argue in favor of a critical line below $T_{SG}$, in our view, other
possibilities can not be ruled out, such as a transition at a lower value
of $T_{SG}$ or even an exponential divergence of the correlation length
at $T=0$.

To distinguish between these scenarios would require a study of still larger
sizes. It will be difficult to go to very much larger sizes without a
better algorithm, since the present study used a quite substantial
amount of computer time. An unfortunate feature of the present algorithm
is that parallel tempering for vector spin models requires a large
number of temperatures.  The large number arises from the temperature
independent specific heat at low temperatures, which, in turn, comes
from from a rather trivial feature: Gaussian spinwave fluctuations.
However, it is difficult to see how to eliminate their effect, and
thereby reduce the number of temperatures.

It is interesting to note that we (and also Campos et
al.\cite{campos:06}) have been able to study larger sizes for the
Heisenberg spin glass ($L=32$) than has ever been done for the Ising
spin glass.  For example,  Katzgraber et al.\cite{katzgraber:06} studied
the Ising spin glass using considerable CPU time, but were still not
able to equilibrate sizes larger than $L=16$ near $T_{SG}$ for the case
of Gaussian interactions.  With $\pm J$ interactions, where ``multispin
coding'' speeds the code up further, they could go up to $L=24$.

That one can equilibrate larger sizes in the Heisenberg case is
surprising bearing in mind that (i) the Heisenberg algorithm is more
complicated and so one sweep takes more CPU time than for the Ising
case, and (ii) the transition temperature is significantly lower for the
Heisenberg case. Noting that the mean field transition temperature is
$T_{SG}^{MF} = \sqrt{z}/m$, where $z$ is the number of neighbors and $m$
the number of spin components, if we take $T_{SG} = 0.145$ for the
Heisenberg case, the ratio $T_{SG}/T_{SG}^{MF}$ is about $0.18$. For the
Ising case, one finds\, e.g. Ref.~\onlinecite{katzgraber:06}, $T_{SG}
\simeq 0.95$ so $T_{SG}/T_{SG}^{MF} \simeq 0.39$ which is more than
twice the corresponding ratio for the Heisenberg spin glass.

The fact that, despite all this, one can study larger sizes in
Heisenberg spin glasses than in Ising spin glasses indicates that
barriers are smaller in the Heisenberg case. Another way of putting this
is that the extra degrees of freedom in the Heisenberg model allow the
system to find a way \textit{round} barriers, which would
have to be \textit{gone over} for the Ising spin glass.

Overall, our results indicate that spins and chiralities behave in a
quite similar manner, and that the lower critical dimension of the
Heisenberg spin glass is close to, and possibly equal to, three.

\begin{acknowledgments}
We acknowledge support
from the National Science Foundation under grant DMR 0337049. We are
also very grateful to the Hierarchical Systems Research Foundation for
a generous allocation of computer time on its Mac G5 cluster.
\end{acknowledgments}

\bibliography{refs,comments}

\begin{thebibliography}{29}
\expandafter\ifx\csname natexlab\endcsname\relax\def\natexlab#1{#1}\fi
\expandafter\ifx\csname bibnamefont\endcsname\relax
  \def\bibnamefont#1{#1}\fi
\expandafter\ifx\csname bibfnamefont\endcsname\relax
  \def\bibfnamefont#1{#1}\fi
\expandafter\ifx\csname citenamefont\endcsname\relax
  \def\citenamefont#1{#1}\fi
\expandafter\ifx\csname url\endcsname\relax
  \def\url#1{\texttt{#1}}\fi
\expandafter\ifx\csname urlprefix\endcsname\relax\def\urlprefix{URL }\fi
\providecommand{\bibinfo}[2]{#2}
\providecommand{\eprint}[2][]{\url{#2}}

\bibitem[{\citenamefont{Ballesteros et~al.}(2000)\citenamefont{Ballesteros,
  Cruz, Fernandez, Martin-Mayor, Pech, Ruiz-Lorenzo, Tarancon, Tellez, Ullod,
  and Ungil}}]{ballesteros:00}
\bibinfo{author}{\bibfnamefont{H.~G.} \bibnamefont{Ballesteros}},
  \bibinfo{author}{\bibfnamefont{A.}~\bibnamefont{Cruz}},
  \bibinfo{author}{\bibfnamefont{L.~A.} \bibnamefont{Fernandez}},
  \bibinfo{author}{\bibfnamefont{V.}~\bibnamefont{Martin-Mayor}},
  \bibinfo{author}{\bibfnamefont{J.}~\bibnamefont{Pech}},
  \bibinfo{author}{\bibfnamefont{J.~J.} \bibnamefont{Ruiz-Lorenzo}},
  \bibinfo{author}{\bibfnamefont{A.}~\bibnamefont{Tarancon}},
  \bibinfo{author}{\bibfnamefont{P.}~\bibnamefont{Tellez}},
  \bibinfo{author}{\bibfnamefont{C.~L.} \bibnamefont{Ullod}}, \bibnamefont{and}
  \bibinfo{author}{\bibfnamefont{C.}~\bibnamefont{Ungil}},
  \emph{\bibinfo{title}{Critical behavior of the three-dimensional {I}sing spin
  glass}}, \bibinfo{journal}{Phys. Rev. B} \textbf{\bibinfo{volume}{62}},
  \bibinfo{pages}{14237} (\bibinfo{year}{2000}),
  \eprint{(arXiv:cond-mat/0006211)}.

\bibitem[{\citenamefont{McMillan}(1985)}]{mcmillan:85b}
\bibinfo{author}{\bibfnamefont{W.~L.} \bibnamefont{McMillan}},
  \emph{\bibinfo{title}{Domain-wall renormalization-group study of the random
  {H}eisenberg model}}, \bibinfo{journal}{Phys. Rev. B}
  \textbf{\bibinfo{volume}{31}}, \bibinfo{pages}{342} (\bibinfo{year}{1985}).

\bibitem[{\citenamefont{Morris et~al.}(1986)\citenamefont{Morris, Colborne,
  Bray, Moore, and Canisius}}]{morris:86}
\bibinfo{author}{\bibfnamefont{B.~M.} \bibnamefont{Morris}},
  \bibinfo{author}{\bibfnamefont{S.~G.} \bibnamefont{Colborne}},
  \bibinfo{author}{\bibfnamefont{A.~J.} \bibnamefont{Bray}},
  \bibinfo{author}{\bibfnamefont{M.~A.} \bibnamefont{Moore}}, \bibnamefont{and}
  \bibinfo{author}{\bibfnamefont{J.}~\bibnamefont{Canisius}},
  \emph{\bibinfo{title}{Zero-temperature critical behaviour of vector spin
  glasses}}, \bibinfo{journal}{J. Phys. C} \textbf{\bibinfo{volume}{19}},
  \bibinfo{pages}{1157} (\bibinfo{year}{1986}).

\bibitem[{\citenamefont{Olive et~al.}(1986)\citenamefont{Olive, Young, and
  Sherrington}}]{olive:86}
\bibinfo{author}{\bibfnamefont{J.~A.} \bibnamefont{Olive}},
  \bibinfo{author}{\bibfnamefont{A.~P.} \bibnamefont{Young}}, \bibnamefont{and}
  \bibinfo{author}{\bibfnamefont{D.}~\bibnamefont{Sherrington}},
  \emph{\bibinfo{title}{A computer simulation of the three dimensional short
  range {H}eisenberg spin glass}}, \bibinfo{journal}{Phys. Rev. B}
  \textbf{\bibinfo{volume}{34}}, \bibinfo{pages}{6341} (\bibinfo{year}{1986}).

\bibitem[{\citenamefont{Villain}(1977)}]{villain:77b}
\bibinfo{author}{\bibfnamefont{J.}~\bibnamefont{Villain}},
  \emph{\bibinfo{title}{Two-level systems in a spin-glass model. {I}. {G}eneral
  formalism and two-dimensional model}}, \bibinfo{journal}{J. Phys. C}
  \textbf{\bibinfo{volume}{10}}, \bibinfo{pages}{4793} (\bibinfo{year}{1977}).

\bibitem[{\citenamefont{Kawamura and Tanemura}(1987)}]{kawamura:87}
\bibinfo{author}{\bibfnamefont{H.}~\bibnamefont{Kawamura}} \bibnamefont{and}
  \bibinfo{author}{\bibfnamefont{M.}~\bibnamefont{Tanemura}},
  \emph{\bibinfo{title}{Chiral order in a two-dimensional {XY} spin glass}},
  \bibinfo{journal}{Phys. Rev. B.} \textbf{\bibinfo{volume}{36}},
  \bibinfo{pages}{7177} (\bibinfo{year}{1987}).

\bibitem[{\citenamefont{Kawamura}(1998)}]{kawamura:98}
\bibinfo{author}{\bibfnamefont{H.}~\bibnamefont{Kawamura}},
  \emph{\bibinfo{title}{Dynamical simulation of of spin-glass and chiral-glass
  orderings in three-dimensional {H}eisenberg spin glasses}},
  \bibinfo{journal}{Phys. Rev. Lett.} \textbf{\bibinfo{volume}{80}},
  \bibinfo{pages}{5421} (\bibinfo{year}{1998}).

\bibitem[{\citenamefont{Hukushima and Kawamura}(2000)}]{hukushima:00}
\bibinfo{author}{\bibfnamefont{K.}~\bibnamefont{Hukushima}} \bibnamefont{and}
  \bibinfo{author}{\bibfnamefont{H.}~\bibnamefont{Kawamura}},
  \emph{\bibinfo{title}{Chiral-glass and replica symmetry breaking of a
  three-dimensional {H}eisenberg spin glass}}, \bibinfo{journal}{Phys. Rev. E}
  \textbf{\bibinfo{volume}{61}}, \bibinfo{pages}{R1008} (\bibinfo{year}{2000}).

\bibitem[{\citenamefont{Hukushima and Kawamura}(2005)}]{hukushima:05}
\bibinfo{author}{\bibfnamefont{K.}~\bibnamefont{Hukushima}} \bibnamefont{and}
  \bibinfo{author}{\bibfnamefont{H.}~\bibnamefont{Kawamura}},
  \emph{\bibinfo{title}{Monte {C}arlo simulations of the phase transition of
  the three-dimensional isotropic {H}eisenberg spin glass}},
  \bibinfo{journal}{Phys. Rev. B} \textbf{\bibinfo{volume}{72}},
  \bibinfo{pages}{144416} (\bibinfo{year}{2005}).

\bibitem[{\citenamefont{Matsubara et~al.}(2001)\citenamefont{Matsubara,
  Shirakura, and Endoh}}]{matsubara:01}
\bibinfo{author}{\bibfnamefont{F.}~\bibnamefont{Matsubara}},
  \bibinfo{author}{\bibfnamefont{T.}~\bibnamefont{Shirakura}},
  \bibnamefont{and} \bibinfo{author}{\bibfnamefont{S.}~\bibnamefont{Endoh}},
  \emph{\bibinfo{title}{Spin and chirality autocorrelation function of a
  {H}eisenberg spin glass model}}, \bibinfo{journal}{Phys. Rev. B}
  \textbf{\bibinfo{volume}{64}}, \bibinfo{pages}{092412}
  (\bibinfo{year}{2001}).

\bibitem[{\citenamefont{Endoh et~al.}(2001)\citenamefont{Endoh, Matsubara, and
  Shirakura}}]{endoh:01}
\bibinfo{author}{\bibfnamefont{S.}~\bibnamefont{Endoh}},
  \bibinfo{author}{\bibfnamefont{F.}~\bibnamefont{Matsubara}},
  \bibnamefont{and}
  \bibinfo{author}{\bibfnamefont{T.}~\bibnamefont{Shirakura}},
  \emph{\bibinfo{title}{Stiffness of the {H}eisenberg spin-glass model at zero-
  and finite-temperatures in three dimensions}}, \bibinfo{journal}{J. Phys.
  Soc. Jpn.} \textbf{\bibinfo{volume}{70}}, \bibinfo{pages}{1543}
  (\bibinfo{year}{2001}).

\bibitem[{\citenamefont{Nakamura and Endoh}(2002)}]{nakamura:02}
\bibinfo{author}{\bibfnamefont{T.}~\bibnamefont{Nakamura}} \bibnamefont{and}
  \bibinfo{author}{\bibfnamefont{S.}~\bibnamefont{Endoh}},
  \emph{\bibinfo{title}{Spin-glass and chiral-glass transitions in a $\pm {J}$
  {H}eisenberg spin-glass model in three dimensions}}, \bibinfo{journal}{J.
  Phys. Soc. Jpn.} \textbf{\bibinfo{volume}{71}}, \bibinfo{pages}{2113}
  (\bibinfo{year}{2002}), \eprint{(arXiv:cond-mat/0110017)}.

\bibitem[{\citenamefont{Picco and Ritort}(2005)}]{picco:05}
\bibinfo{author}{\bibfnamefont{M.}~\bibnamefont{Picco}} \bibnamefont{and}
  \bibinfo{author}{\bibfnamefont{F.}~\bibnamefont{Ritort}},
  \emph{\bibinfo{title}{Dynamical ac study of the critical behavior in
  {H}eisenberg spin glasses}}, \bibinfo{journal}{Phys. Rev. B}
  \textbf{\bibinfo{volume}{71}}, \bibinfo{pages}{100406(R)}
  (\bibinfo{year}{2005}).

\bibitem[{\citenamefont{Lee and Young}(2003{\natexlab{a}})}]{leeLW:03}
\bibinfo{author}{\bibfnamefont{L.~W.} \bibnamefont{Lee}} \bibnamefont{and}
  \bibinfo{author}{\bibfnamefont{A.~P.} \bibnamefont{Young}},
  \emph{\bibinfo{title}{Single spin- and chiral-glass transition in vector spin
  glasses in three-dimensions}}, \bibinfo{journal}{Phys. Rev. Lett.}
  \textbf{\bibinfo{volume}{90}}, \bibinfo{pages}{227203}
  (\bibinfo{year}{2003}{\natexlab{a}}), \bibinfo{note}{(referred to as LY)},
  \eprint{(arXiv:cond-mat/0302371)}.

\bibitem[{\citenamefont{Campos et~al.}(2006)\citenamefont{Campos, Cotallo-Aban,
  Martin-Mayor, Perez-Gaviro, and Tarancon}}]{campos:06}
\bibinfo{author}{\bibfnamefont{I.}~\bibnamefont{Campos}},
  \bibinfo{author}{\bibfnamefont{M.}~\bibnamefont{Cotallo-Aban}},
  \bibinfo{author}{\bibfnamefont{V.}~\bibnamefont{Martin-Mayor}},
  \bibinfo{author}{\bibfnamefont{S.}~\bibnamefont{Perez-Gaviro}},
  \bibnamefont{and} \bibinfo{author}{\bibfnamefont{A.}~\bibnamefont{Tarancon}},
  \emph{\bibinfo{title}{Spin-glass transition of the three-dimensional
  {H}eisenberg spin glass}}, \bibinfo{journal}{Phys. Rev. Lett.}
  \textbf{\bibinfo{volume}{97}}, \bibinfo{pages}{217204}
  (\bibinfo{year}{2006}).

\bibitem[{\citenamefont{Kosterlitz and Thouless}(1973)}]{kosterlitz:73}
\bibinfo{author}{\bibfnamefont{J.~M.} \bibnamefont{Kosterlitz}}
  \bibnamefont{and} \bibinfo{author}{\bibfnamefont{D.~J.}
  \bibnamefont{Thouless}}, \emph{\bibinfo{title}{Ordering, metastability and
  phase transitions in two-dimensional systems}}, \bibinfo{journal}{J. Phys. C}
  \textbf{\bibinfo{volume}{6}}, \bibinfo{pages}{1181} (\bibinfo{year}{1973}).

\bibitem[{\citenamefont{Berezinskii}(1970)}]{berezinskii:70}
\bibinfo{author}{\bibfnamefont{V.~L.} \bibnamefont{Berezinskii}},
  \emph{\bibinfo{title}{Destruction of long range order in one-dimensional and
  two-dimensional systems having a continuous symmetry group; {I} classical
  systems}}, \bibinfo{journal}{Sov. Phys. JETP} \textbf{\bibinfo{volume}{32}},
  \bibinfo{pages}{493} (\bibinfo{year}{1970}).

\bibitem[{\citenamefont{Campbell and Kawamura}(2007)}]{campbell:07}
\bibinfo{author}{\bibfnamefont{I.~A.} \bibnamefont{Campbell}} \bibnamefont{and}
  \bibinfo{author}{\bibfnamefont{H.}~\bibnamefont{Kawamura}},
  \emph{\bibinfo{title}{Comment on ``spin-glass transition of the
  three-dimensional {H}eisenberg spin glass"}} (\bibinfo{year}{2007}),
  \eprint{(arXiv:cond-mat/0703369)}.

\bibitem[{\citenamefont{Hukushima and Nemoto}(1996)}]{hukushima:96}
\bibinfo{author}{\bibfnamefont{K.}~\bibnamefont{Hukushima}} \bibnamefont{and}
  \bibinfo{author}{\bibfnamefont{K.}~\bibnamefont{Nemoto}},
  \emph{\bibinfo{title}{Exchange {M}onte {C}arlo method and application to spin
  glass simulations}}, \bibinfo{journal}{J. Phys. Soc. Japan}
  \textbf{\bibinfo{volume}{65}}, \bibinfo{pages}{1604} (\bibinfo{year}{1996}).

\bibitem[{\citenamefont{Marinari}(1998)}]{marinari:98b}
\bibinfo{author}{\bibfnamefont{E.}~\bibnamefont{Marinari}},
  \emph{\bibinfo{title}{Optimized {M}onte {C}arlo methods}}, in
  \emph{\bibinfo{booktitle}{Advances in Computer Simulation}}, edited by
  \bibinfo{editor}{\bibfnamefont{J.}~\bibnamefont{Kert\'esz}} \bibnamefont{and}
  \bibinfo{editor}{\bibfnamefont{I.}~\bibnamefont{Kondor}}
  (\bibinfo{publisher}{Springer-Verlag}, \bibinfo{year}{1998}),
  p.~\bibinfo{pages}{50}, \eprint{(arXiv:cond-mat/9612010)}.

\bibitem[{\citenamefont{Palassini and Caracciolo}(1999)}]{palassini:99b}
\bibinfo{author}{\bibfnamefont{M.}~\bibnamefont{Palassini}} \bibnamefont{and}
  \bibinfo{author}{\bibfnamefont{S.}~\bibnamefont{Caracciolo}},
  \emph{\bibinfo{title}{Universal finite size scaling functions in the 3d
  {I}sing spin glass}}, \bibinfo{journal}{Phys. Rev. Lett.}
  \textbf{\bibinfo{volume}{82}}, \bibinfo{pages}{5128} (\bibinfo{year}{1999}),
  \eprint{(arXiv:cond-mat/9904246)}.

\bibitem[{\citenamefont{Alonso et~al.}(1986)\citenamefont{Alonso,
  A.~Taranc\'on, Ballesteros, Fern\'andez, Mart\'in-Mayor, and Mu\~noz
  Sudupe}}]{alonso:96}
\bibinfo{author}{\bibfnamefont{J.}~\bibnamefont{Alonso}},
  \bibinfo{author}{\bibfnamefont{A.}~\bibnamefont{A.~Taranc\'on}},
  \bibinfo{author}{\bibfnamefont{H.}~\bibnamefont{Ballesteros}},
  \bibinfo{author}{\bibfnamefont{L.}~\bibnamefont{Fern\'andez}},
  \bibinfo{author}{\bibfnamefont{V.}~\bibnamefont{Mart\'in-Mayor}},
  \bibnamefont{and} \bibinfo{author}{\bibfnamefont{A.}~\bibnamefont{Mu\~noz
  Sudupe}}, \emph{\bibinfo{title}{Monte {C}arlo study of {O}(3)
  antiferromagnetic models in three dimensions}}, \bibinfo{journal}{Phys. Rev.
  B} \textbf{\bibinfo{volume}{53}}, \bibinfo{pages}{2537}
  (\bibinfo{year}{1986}).

\bibitem[{com()}]{comment1}
\bibinfo{note}{See, for example, Secs.~3.2.2 and 3.4.3 of
  Ref.~\onlinecite{berg:04}. In our opinion, the factor of $\sin\alpha^r
  \cos\beta^r \cos\phi$ in the second line of Eq.~(3.62) in that reference
  should be $\sin\alpha^r \sin\beta^r \cos\phi$.}

\bibitem[{\citenamefont{Katzgraber et~al.}(2001)\citenamefont{Katzgraber,
  Palassini, and Young}}]{katzgraber:01}
\bibinfo{author}{\bibfnamefont{H.~G.} \bibnamefont{Katzgraber}},
  \bibinfo{author}{\bibfnamefont{M.}~\bibnamefont{Palassini}},
  \bibnamefont{and} \bibinfo{author}{\bibfnamefont{A.~P.} \bibnamefont{Young}},
  \emph{\bibinfo{title}{{M}onte {C}arlo simulations of spin glasses at low
  temperatures}}, \bibinfo{journal}{Phys. Rev. B}
  \textbf{\bibinfo{volume}{63}}, \bibinfo{pages}{184422}
  (\bibinfo{year}{2001}), \eprint{(arXiv:cond-mat/0108320)}.

\bibitem[{\citenamefont{Katzgraber and Young}(2002)}]{katzgraber:01c}
\bibinfo{author}{\bibfnamefont{H.~G.} \bibnamefont{Katzgraber}}
  \bibnamefont{and} \bibinfo{author}{\bibfnamefont{A.~P.} \bibnamefont{Young}},
  \emph{\bibinfo{title}{{M}onte {C}arlo simulations of the four dimensional
  {XY} spin-glass at low temperatures}}, \bibinfo{journal}{Phys. Rev. B}
  \textbf{\bibinfo{volume}{65}}, \bibinfo{pages}{214401}
  (\bibinfo{year}{2002}), \eprint{(arXiv:cond-mat/0108320)}.

\bibitem[{\citenamefont{Palassini and Young}(1999)}]{palassini:99}
\bibinfo{author}{\bibfnamefont{M.}~\bibnamefont{Palassini}} \bibnamefont{and}
  \bibinfo{author}{\bibfnamefont{A.~P.} \bibnamefont{Young}},
  \emph{\bibinfo{title}{Triviality of the ground state structure in {I}sing
  spin glasses}}, \bibinfo{journal}{Phys. Rev. Lett.}
  \textbf{\bibinfo{volume}{83}}, \bibinfo{pages}{5126} (\bibinfo{year}{1999}),
  \eprint{(arXiv:cond-mat/9906323)}.

\bibitem[{\citenamefont{Lee and Young}(2003{\natexlab{b}})}]{lee:03}
\bibinfo{author}{\bibfnamefont{L.~W.} \bibnamefont{Lee}} \bibnamefont{and}
  \bibinfo{author}{\bibfnamefont{A.~P.} \bibnamefont{Young}},
  \emph{\bibinfo{title}{Single spin- and chiral-glass transition in vector spin
  glasses in three-dimensions}}, \bibinfo{journal}{Phys. Rev. Lett.}
  \textbf{\bibinfo{volume}{90}}, \bibinfo{pages}{227203}
  (\bibinfo{year}{2003}{\natexlab{b}}), \eprint{(arXiv:cond-mat/0302371)}.

\bibitem[{\citenamefont{Katzgraber et~al.}(2006)\citenamefont{Katzgraber,
  K\"orner, and Young}}]{katzgraber:06}
\bibinfo{author}{\bibfnamefont{H.~G.} \bibnamefont{Katzgraber}},
  \bibinfo{author}{\bibfnamefont{M.}~\bibnamefont{K\"orner}}, \bibnamefont{and}
  \bibinfo{author}{\bibfnamefont{A.~P.} \bibnamefont{Young}},
  \emph{\bibinfo{title}{Detailed study of universality in three-dimensional
  {I}sing spin glasses}}, \bibinfo{journal}{Phys. Rev. B}
  \textbf{\bibinfo{volume}{73}}, \bibinfo{pages}{224432}
  (\bibinfo{year}{2006}), \eprint{(arXiv:cond-mat/0602212)}.

\bibitem[{\citenamefont{Berg}(2004)}]{berg:04}
\bibinfo{author}{\bibfnamefont{B.~A.} \bibnamefont{Berg}},
  \emph{\bibinfo{title}{Markov Chain {M}onte {C}arlo Simulations and Their
  Statistical Aplications}} (\bibinfo{publisher}{World Scientific},
  \bibinfo{address}{Singapore}, \bibinfo{year}{2004}).

\end{thebibliography}

\end{document}